\newcommand\bm[1]{\mbox{\boldmath$#1$}}

\def\sqr#1#2{{\vcenter{\vbox{\hrule height.#2pt
	\hbox{\vrule width.#2pt height#1pt \kern#1pt
	\vrule width.#2pt}
	\hrule height.#2pt}}}}
\def\square{\mathchoice\sqr65\sqr65\sqr44\sqr34}

\documentclass[12pt,a4paper]{article}

\begin{document}

\title{An approximate global solution  of Einstein's \\ equations 
for a finite body}
\author{J. A. Cabezas${}^1$ J.\ Mart\'{\i}n${}^2$,
A.\ Molina${}^3$ and E.\ Ruiz${}^2$\\[.5ex]
${}^1$\emph{Departamento de Ingenier\'\i{}a Mec\'anica},\\
${}^2$\emph{Departamento de F\'\i sica Fundamental},
\\
\emph{Facultad de Ciencias, Universidad de Salamanca},
\\
\emph{Plaza de la Merced s/n, 37008 Salamanca, Spain}.
\\
${}^3$\emph{Departament\ de F\'\i sica Fonamental},
\\
\emph{Universitat de Barcelona},
\\
\emph{Diagonal 647, Barcelona 08028}.
}
\maketitle

\date{}

\begin{abstract}
We obtain an approximate global stationary and axisymmetric solution
of Einstein's equations which can be considered as a simple star model: a self-gravitating perfect fluid ball with constant mass density rotating in rigid
motion. Using the post-Min\-kows\-kian formalism (weak-field approximation) and considering rotation as a perturbation 
(slow-rotation
approximation), we find approximate interior and exterior 
(asymptotically flat)
solutions to this problem in harmonic and quo-harmonic
coordinates. In both cases, interior and exterior solutions are matched, in the sense 
of Lichnerowicz, on the
surface of zero pressure to obtain a global solution. The  resulting metric depends on three 
arbitrary constants: mass density,
rotational velocity and  the star radius at the non-rotation limit. 
The mass,
angular momentum,  quadrupole moment and other constants of the exterior metric are determined by these three
parameters. It is easy  to show that this type of fluid cannot be a 
source of the Kerr metric.
\end{abstract}

keywords: relativistic astrophysics, two post-minkowskian approximation, harmonic coordinates, rotating stars

\section{Introduction}

Rotating stars have been an active field of research in General Relativity in the last sixty years (see
\cite{Krasinski1, Krasinski2, Krasinski3,Krasinski4, Stergioulas} and references there in). Nevertheless, until the discovery of neutron stars, this research was mainly theoretical since the relativistic effects of rotation are negligible for normal stars. The high angular velocit and small radius of neutron stars have stimulated new efforts to understand the
gravitational fields outside and inside a rotating star in a non-Newtonian framework.

Only a few exact solutions of the Einstein equations can describe the gravitational field inside a rotating star, i. e., the
gravitational field of an axisymmetric rotating perfect fluid (see  \cite {Senovilla} for a brief review and  \cite{ Mars1, Mars2}). So far it has not been
possible to match any of them to an asymptotically flat vacuum solution, the  gravitational field outside the star, to obtain a global star model similar to
those provided by Newtonian theory. This lack of examples does not mean the absence of theoretical results to this problem even in full nonlinear General
Relativity. For instance, axisymmetric matching conditions have been largely revised and used to demonstrate the uniqueness of the exterior gravitational field to an isolated
body
\cite{Marc-Seno}. Nevertheless the most relevant outcomes result from approximation methods and numerical computation. Both techniques have been applied to
obtain useful descriptions of the gravitational field of a star obeying different equations of state (relativistic polytropes, hadronic fluids, etc\dots). They have also been used to study other related problems such as stability, normal modes of vibration and generation of gravitational waves (all these topics are extensively reviewed in
\cite{Stergioulas}).

We are interested in analytic
approximation methods. Obviously, models that can be handled with these methods have to be simpler than those amenable to numerical methods. Analytic approaches provide a better understanding of the basic features of relativistic rotating stars, so they be of some help to study
more interesting and realistic models which deserve more powerful methods.

On these bases we introduce a new analytic approximation scheme and apply it to search for an approximate global solution to the gravitational field of a
fluid with a very simple equation of state.
We aim to establish  an approximate solution of the Einstein equations which describes the gravitational field
inside a ball of perfect fluid with constant mass density rotating in a rigid motion, and to match it, on the
zero pressure surface, to an asymptotically flat approximate solution of the vacuum Einstein equations. We solve this problem in harmonic coordinates and
quo-harmonic coordinates (harmonic coordinates in the sense of the quotient metric). In both cases, we require the metric to be of class $C^1$ on the matching
surface, that is the metric should satisfy the Lichnerowicz matching conditions \cite{Lichnerowicz}.

Two classes of analytic approximation methods are described in the literature. First, the post-Newtonian
approximation scheme, which computes perturbations to well-known
Newtonian solutions such as the McLaurin ellipsoids
\cite{Chandrasekhar1, Chandrasekhar2}; second, the slow rotation approach, which
calculates perturbations to spherically symmetric solutions of the
Einstein equations \cite{Hartle}. The approximation scheme we propose is an intermediate way: we implement a slow rotation approximation
on a post-Minkowskian algorithm  (a first attempt can be found in
\cite{Cabezas}). We introduce two dimensionless
parameters. One, $\lambda$, measures the strength of the
gravitational field, the other, $\Omega$, measures the speed of rotation of the fluid.
We also make some assumptions about how all the constants that appear in
the post-Minkowskian metric depend on these two parameters. So we obtain
an expansion of the metric in a double power series of $\lambda$ and
$\Omega$. Obviously, if there is no rotation ($\Omega=0$), we are faced
to the post-Minkowskian perturbation to the Newtonian gravitational
field of a spherically symmetric mass distribution. On the other hand,
Newtonian deformation of the source due to rotation is included in first
order $\lambda$ terms up to some order in the rotation parameter. In other words, there is no exact Newtonian rotational 
model underlying our
approach but only an approximate one. Nevertheless, we use these linear terms
to compute non-linear contributions of rotation to the gravitational
field.

We only evaluate metric terms of order less than or equal to
$\Omega^3$ and no greater than $\lambda^2$. However, since the algorithm
is implemented by an algebraic computational program, our
results  can easily be enhanced, if so desired, by going farther in
the approximation scheme. The values taken by the two
control parameters for a definite system tell us which terms in the
double expansion of the gravitational field are relevant to that particular
problem.

In section $2$ we list the properties we impose on the space-time and the energy-momentum tensor and we define the notation. Moreover we briefly review the post-Minkowskian approximation we use in the following sections.

In section $3$ we obtain the first order post-Minkowskian metric up to the second order in the rotation parameter. We solve the linearized
Einstein equations outside and inside the mass distribution. We discuss the dependence of the free constants in the linear solution on $\lambda$ and $\Omega$.
We use that information to cancel all terms in the solution of order higher than $\Omega^3$. Finally we match those simplified exterior and interior
solutions on the appropriate surface up to this order of approximation. Consequently we derive expressions  for all the surviving free constants of the exterior
and interior gravitational fields as functions of the parameters of the source, its mass density, mean radius and angular velocity. A global system of harmonic coordinates is assumed.

In section $4$ we work up the second-order
approximation. 

The role of frames of reference in General Relativity \cite{Bel} leads us to focus our attention on the quo-harmonic coordinates. In section $5$ we show
that our scheme can also be successfully applied by imposing on the metric this unusual but interesting coordinate condition.

\section{Preliminaries: notation and other topics}

\subsection{Metric}

We require space-time to be a stationary and
axisymmetric Riemannian manifold admitting a global system of 
spherical-like
coordinates $\{t,r,\theta,\varphi\}$ which verifies the following properties:

\begin{enumerate}

\item Coordinates are adapted to the space-time symmetry, $\bm\xi=\partial_t$ and
$\bm\zeta=\partial_\varphi$ are respectively the timelike 
and spacelike
Killing vectors, so that the metric components do not
depend on coordinates
$t$ or $\varphi$.

\item Coordinates $\{r,\theta\}$ parametrize two dimensional surfaces
orthogonal to the orbits of the symmetry group, that is the metric
tensor has Papapetrou structure,
\begin{eqnarray}
&\bm{g} = \gamma_{tt}\,\bm{\omega}^t{\otimes\,}\bm{\omega}^t
+\gamma_{t\varphi}(\bm{\omega}^t{\otimes\,}\bm{\omega}^\varphi+\bm{\omega}^\varphi{\otimes\,}\bm{\omega}^t)+
\gamma_{\varphi\varphi}\,\bm{\omega}^\varphi{\otimes\,}\bm{\omega}^\varphi
\nonumber\\
&\quad +\,\,\gamma_{rr}\,\bm{\omega}^r{\otimes\,}\bm{\omega}^r+
\gamma_{r\theta}(\bm{\omega}^r{\otimes\,}\bm{\omega}^\theta+\bm{\omega}^\theta{\otimes\,}\bm{\omega}^r) 
+\gamma_{\theta\theta}\,\bm{\omega}^\theta{\otimes\,}\bm{\omega}^\theta\,,
\label{eqmetrica}
\end{eqnarray}
where
$\bm{\omega}^t=dt$, $\bm{\omega}^r=dr$, $\bm{\omega}^\theta=r\,d\theta$, $\bm{\omega}^\varphi=r\sin\theta\,d\varphi$ 
is the Euclidean orthonormal cobasis associated to these coordinates.

\item Coordinates $\{t,\,x=r\sin\theta\cos\varphi,\,y=r\sin\theta\sin\varphi,\,
z=\cos\theta\}$ associated with the spherical-like 
coordinates
 are Cartesian coordinates at spacelike infinity, that is the metric in these
coordinates tends to the Min\-kows\-ky metric in standard Cartesian 
coordinates for large values of the coordinate $r$.

\end{enumerate}

All these properties are compatible with the two classes of coordinates we use in this paper, mainly harmonic and
quo-harmonic coordinates. Time coordinate $t$ is always harmonic under
assumptions $1$ and $2$, then we have only to check that spatial
coordinates $\{x,y,z\}$ verify the coordinate condition we choose.

\subsection{Energy-momentum tensor}
We assume that the source of the gravitational field is a perfect fluid,
\begin{equation}
\bm{T} = \left(\mu + p\right)\bm{u}\otimes\bm{u}+p\,\bm{g}\,,
\label{eqenermom}
\end{equation}
whose density $\mu$ is a constant and whose pressure $p$ depends only on $r$ and
$\theta$ coordinates. We also assume that the fluid has no convective motion, so its velocity
$\bm{u}$ lies on the plane spanned by the two Killing vectors,
\begin{equation}
\bm{u} = \psi\left(\bm\xi + \omega\,\bm\zeta\right)\,,
\label{velocidad}
\end{equation}
where
\begin{equation}
\psi \equiv \left[-\left(\gamma_{tt}+2\omega\,\gamma_{t\varphi}\,r\sin\theta+\omega^2
\,\gamma_{\varphi\varphi}\,r^2\sin^2\theta\right)\right]^{-\frac12}
\label{eqnorma}
\end{equation}
is a normalization factor.

Let us consider Euler equations for the fluid (or, what is equivalent, the energy-momentum tensor conservation law)
to gain information about the matching surface. Assuming the fluid rotates rigidly as we do, $\omega={\rm 
constant}$, the equations read \cite{Boyer}
\begin{equation}
\partial_a p = (\mu + p)\partial_a\ln\psi
\qquad (a,b,\dots = r\,,\theta)\,.
\end{equation}
Since $\mu$ depends neither on $r$ nor on $\theta$, we can integrate them 
to give a very simple expression for the pressure in terms of the normalization factor
\begin{equation}
\qquad p = \mu\left(\frac{\psi}{\psi_\Sigma}-1\right)\,,
\label{eqpressio1}
\end{equation}
which in turn leads to the following implicit equation for the matching surface 
\begin{equation}
p=0\quad \Longleftrightarrow \quad\psi=\psi_\Sigma\,,
\label{eqsuperficie}
\end{equation}
here $\psi_\Sigma$ is an arbitrary constant.

These last two equations, (\ref{eqpressio1}) and (\ref{eqsuperficie}),  play an important role in our scheme. We use them to derive approximate expressions for the
pressure and the matching surface in a coherent way with the expansion of the metric we propose.

\subsection{Einstein's equations: Post-Minkowskian appro\-xi\-ma\-tion}
The Einstein equations for any energy-momentum tensor read
\begin{equation}
R_{\alpha\beta} = 8\pi\left(T_{\alpha\beta}-\frac12 g_{\alpha\beta}T\right) \equiv 8\pi  t_{\alpha\beta}\,.
\label{einsteineq}
\end{equation}
The deviation of the metric tensor from the Minkowski metric,
\begin{equation}
h_{\alpha\beta}\equiv  g_{\alpha\beta}-\eta_{\alpha\beta}\,,
\end{equation}
in a system of Cartesian-like coordinates, where $(\eta_{\alpha\beta})={\rm diag}(-1,1,1,1)$, provides the natural object to work out a post-Minkowskian
approximation. If we substitute the above expression for the metric into equation (\ref{einsteineq}) and neglect all the quadratic and higher terms in
$h_{\alpha\beta}$, we obtain the well known linear approximation to the  Einstein equations \cite{MTW}, but if we do not neglect these  terms we can use equation
(\ref{einsteineq}) to  generate a set of successive approximations to the exact solution $g_{\alpha\beta}$. This approximation method is usually  implemented in
harmonic coordinates,
\begin{equation}
g^{\lambda\mu}\Gamma^\rho_{\lambda\mu} = 0\,,
\label{harmoniceq}
\end{equation}
although another type of coordinates could be used (for instance, those
we shall use in section $5$, i. e., quo-harmonic 
coordinates).

Using the metric deviation, the Ricci tensor can be written as
\begin{eqnarray}
&&R_{\alpha\beta} = -\frac12\square h_{\alpha\beta} +
\frac12\partial_\alpha\!\left[\partial^\rho\left(h_{\rho\beta}-
\frac12\eta_{\rho\beta}\,h\right)\right]
\nonumber\\
&&\quad\quad
+\,\,\frac12\partial_\beta\!\left[\partial^\rho\!\left(h_{\rho\alpha}-
\frac12\eta_{\rho\alpha}\,h\right)\right] +N_{\alpha\beta}\,,
\end{eqnarray}
where $\square=\eta^{\alpha\beta}\partial_\alpha\partial_\beta\,$, 
$h=\eta^{\alpha\beta}h_{\alpha\beta}\,$,
$\partial^\rho=\eta^{\rho\sigma}\partial_\sigma\,$ and $N_{\alpha\beta}$ collects
quadratic and higher terms in $h_{\alpha\beta}$. Analogously the harmonic
condition (\ref{harmoniceq}) takes the form
\begin{equation}
\eta_{\rho\alpha}g^{\lambda\mu}\Gamma^\rho_{\lambda\mu}=
\partial^\rho\left(h_{\rho\alpha}-\frac12\eta_{\rho\alpha}\,h\right) +H_\alpha=0\,,
\end{equation}
where  $H_\alpha$ also takes account of non linear terms. These equations
reduce to
\begin{eqnarray}
&&\triangle h_{\alpha\beta} = 2\left(N_{\alpha\beta}-8\pi t_{\alpha\beta} -
\partial_{(\alpha}H_{\beta)}\right)\equiv S_{\alpha\beta}\,,\nonumber\\
[.6ex]
&&\partial^j\left(h_{j\alpha}-\frac12\eta_{j\alpha}\,h\right) =
-H_\alpha\,,
\label{equiteracio}
\end{eqnarray}
when we look for stationary metrics, $\partial_0 h_{\alpha\beta}=0$.

Let us remark that something must be said about how the energy-momen\-tum tensor depends on the metric in order to obtain a true perturbation scheme from
the above equations. Such information is given by formulas (\ref{eqenermom}), (\ref{velocidad}), (\ref{eqnorma}) and (\ref{eqpressio1}) for the type of fluids we are considering in this paper if we
identify the Cartesian-like coordinates $\{t,x,y,z\}$ of our metric (\ref{eqmetrica}) with the coordinates
 $\{x^{\alpha}\}$ used in the post-Minkowskian algorithm.

\section{First order metric in harmonic coordinates}
\subsection{Linear exterior solution}
In the absence of matter, $t_{\alpha\beta}=0$, equations (\ref{equiteracio}) becomes linear if all the quadratic and higher order terms in the metric deviation
$h_{\alpha\beta}$ are neglected, that is $N_{\alpha\beta}=H_\alpha=0$,
\begin{eqnarray}
&&\triangle h_{\alpha\beta} = 0\,,\nonumber\\
[.6ex]
&&\partial^k (h_{k\mu} -\frac12
h\,\eta_{k\mu}\,) = 0\,.
\label{eqhomog}
\end{eqnarray}
The general axisymmetric solution of these homogeneous equations which is compatible with the Papapetrou structure we have assumed for the metric
[see formula (\ref{eqmetrica})] and has a regular behavior at infinity is well known \cite{Thorne1, Elba}. It can be written as follows:
\begin{eqnarray}
&&\bm{h}=2\sum_{n=0}^\infty\frac{M_n}{r^{n+1}}\left(\bm{T}_n+\bm{D}_n\right)
+2\sum_{n=1}^\infty\frac{J_n}{r^{n+1}}\,\bm{Z}_n
\nonumber\\
&&\quad\,\,\,\,+\,\sum_{n=1}^\infty\frac{A_n}{r^{n+1}}\,\bm{E}_n
+\sum_{n=2}^\infty\frac{B_n}{r^{n+1}}\,\bm{F}_{n}\,,
\label{eqsolinexthar}
\end{eqnarray}
where
\begin{eqnarray}
&&\bm{T}_n\, \equiv P_n(\cos\theta)\,\bm{\omega}^t\otimes\,\bm{\omega}^t \quad (n\geq 0)\,,
\nonumber\\
&&\bm{D}_n \equiv P_n(\cos\theta)\,\delta_{ij}dx^i{\otimes\,}dx^j \quad (n\geq 0)\,,
\nonumber\\
&&\bm{Z}_n \,\equiv
P_n^1(\cos\theta)\,(\bm{\omega}^t\otimes\bm{\omega}^\varphi+\bm{\omega}^\varphi\otimes\bm{\omega}^t)
\quad (n\geq 1)\,,
\label{base1}
\end{eqnarray}
are spherical harmonic tensors and
\begin{eqnarray}
&&\bm{E}_n \equiv -\frac23n\,\bm{D}_n +\frac23n\,\bm{H}^0_n +\bm{H}^1_n \quad (n\geq 1)\,,
\nonumber\\
&&\bm{F}_n \equiv \frac12n(n-1)\,\bm{H}^0_n+(n-1)\,\bm{H}^1_n -\frac12\,\bm{H}^2_n \quad (n\geq 2)
\label{basext}
\end{eqnarray}
are two suitable combinations of $\bm{D}_n$ and these other three spherical harmonic tensors
\begin{eqnarray}
&&\bm{H}^0_n \equiv P_n(\cos\theta)\,(\delta_{ij}- 3e_i  e_j)dx^i{\otimes\,}dx^j \quad (n\geq 0)\,,
\nonumber \\
&&\bm{H}^1_n \equiv P_n^1(\cos\theta)\,(k_ie_j + k_je_i)dx^i{\otimes\,} dx^j \quad (n\geq 1)\,,
\nonumber \\
&&\bm{H}^2_n \equiv P_n^2(\cos\theta)\,(k_ik_j -m_im_j)dx^i{\otimes\,} dx^j \quad (n\geq 2)
\label{base2}
\end{eqnarray}
$k_i$, $e_i$ and $m_i$ stand for Euclidean unit vectors of standard cylindrical coordinates, 
$d\rho=k_i\,dx^i$, $dz = e_i\,dx^i$, $\rho\,d\varphi = m_i\,dx^i$ (this is the set  of spherical harmonic tensors we use to write covariant tensors of rank-2 in
this paper); $M_n$ and $J_n$ are the multipole moments of Thorne \cite{Thorne1} or Geroch-Hansen \cite{Geroch} (both definitions are equivalent
\cite{Gursel}) and
$A_n$ and
$B_n$ are other constants that can be related to ``stress moments'' of the source \cite{Ruiz}. These last two sets of constants unlike $M_n$ and $J_n$ are not
intrinsic,  but they  are necessary to solve the Lichnerowicz matching problem.

$\bm{F}_2$ has spherical symmetry (therefore it
must be included in the spherical symmetric linear solution besides the
mass monopole term). This is easy to verify from the  following
alternative expression (indices are raised with the Euclidean metric)
\begin{equation}
F_2^{ij} \equiv H_2^{0\,\, ij}+H_2^{1\,\,ij}-\frac12 H_2^{2\,\,ij}=\delta^{ij}-3n^in^j\,,
\end{equation}
where $n^i$ is the Euclidean unit radial vector of standard spherical
coordinates.

We want the metric to have equatorial symmetry, so we set
$M_{2n+1}=0\,$, $J_{2n}=0\,$, $A_{2n+1}=0$ and $B_{2n+1}=0$ in equation (\ref{eqsolinexthar}).

\subsection{Multipole moments and control parameters}
This section is the key to understandig the current analysis. We  expect  all those constants of the exterior metric (\ref{eqsolinexthar}) to be functions  of  the parameters
included in the energy-momentum tensor of the fluid; i. e., the mass density $\mu$, the angular velocity $\omega$ and the value  $\psi_\Sigma$  of the
normalization factor on the zero pressure surface. Before making  any  assumption on the dependence of those exterior constants on these parameters, let us 
introduce another set of related parameters which will be more suitable to this end.

First let us denote by $r_0$ the typical size or mean radius of  the  source, it can also be viewed as the radius of the sphere towards which the matching  surface 
would tend at the non-rotating limit. Therefore $m = \frac43\pi\mu r_0^3$ may  provide a good estimate of the total mass of the source and the dimensionless 
parameter $\lambda = m/r_0$ (the ratio of Schwarzschild's radius to the mean radius of the source) may give a measure of the strength of the gravitational field.
We shall assume  that this parameter is small and use it to control the post-Minkowskian  expansion  of the metric, in the sense that none of the terms of the
linear exterior solution can  be of order $\lambda^2$ or greater.

Now let us take a look to the implicit equation of the matching surface. We expect that expression (\ref{eqsuperficie}) defines a slightly deformed sphere, $r\approx r_0$, at the lowest order of
approximation. Nevertheless, this is not compatible with the post-Minkowskian approach, because it imposes at this level $\gamma_{tt}\approx -1+O(\lambda)$,
$\gamma_{t\varphi}\approx O(\lambda)$ and $\gamma_{\varphi\varphi}\approx 1+O(\lambda)$ what leads to a cylinder, $\rho={\rm constant}$, not to a sphere, unless we
assume that $\omega^2$ is at least of order $\lambda$. That suggests that we should introduce another dimensionless parameter $\Omega$, such that
$\Omega^2={r_0^2\omega^2/\lambda}$, instead of the apparently more physical parameter $r_0\,\omega$. This new parameter can be seen as an estimate of the
ratio  of centrifugal to gravitational forces or rotational to gravitational energies on the surface of the source \cite{Thorne2}. At all extends $\Omega$
is a parameter that takes account of the deviation of the solution from spherical symmetry due
to  rotation.

Once we have decided the parameters we are going to use in our 
approach, we have to elucidate how the constants in (\ref{eqsolinexthar}) depend on them. Our proposal can be summarized in the following set of substitutions (we use the same notation for old and new constants):
\begin{eqnarray}
M_n &\rightarrow& m r_0^n M_n = \lambda r_0^{n+1} M_n \ \rightarrow\ \lambda\Omega^n r_0^{n+1} M_n\,,
\nonumber\\
J_n &\rightarrow& m\omega r_0^{n+1} J_n = \lambda^{3/2}\Omega r_0^{n+1} J_n \ \rightarrow\ \lambda^{3/2}\Omega^n r_0^{n+1} J_n\,,
\nonumber\\
A_n &\rightarrow& \lambda r_0^{n+1}\,A_n \ \rightarrow\ \lambda\Omega^n r_0^{n+1} A_n\,,
\nonumber\\
B_n &\rightarrow&  \lambda r_0^{n+1} B_n \ \rightarrow\ \lambda\Omega^{n-2} r_0^{n+1} B_n\,.
\label{substitution}
\end{eqnarray}
The first step drives multipole moments to a dimensionless form. An 
extra factor $\Omega$ is included in all $J_n$ multipole moments due to their clear  dynamical character, they should vanish if there is no rotational motion of the fluid.  This feature has also been taken into account in the second step, since all multipole moments except
$M_0$ and $B_2$ (as it was pointed out at the end of the preceding section)
should also vanish if the solution has spherical symmetry. The leading  power of $\Omega$ we assign to each constant in this second set of substitutions reflects just the order  of the spherical
harmonic polynomial to which it is related. This is partly 
explained by the structure of the mass multipole moments of the MacLaurin ellipsoids: the leading term in the expansion in powers of $\Omega$ of these quantities shows the same dependence on this parameter \cite{Cabezas}.

Substituting (\ref{substitution}) into expression (\ref{eqsolinexthar}) and neglecting terms of order  equal or greater than $\Omega^4$ (slow rotation) we arrive at the following approximate expression for the exterior metric
\begin{eqnarray}
&&\bm{g}_{\rm ext}
\approx\left(-1+2\lambda\frac{M_0}{\eta}\right)\bm{T}_0
+\left(1+2\lambda\frac{M_0}{\eta}\right)\bm{D}_0
+\lambda\frac{B_2}{\eta^3}\,\bm{F}_2
\nonumber\\
&&\qquad
+\,\,2\lambda\Omega^2\frac{M_2}{\eta^3}\left(\bm{T}_2+\bm{D}_2\right)
+\lambda\Omega^2\frac{A_2}{\eta^3}\,\bm{E}_2 
+\lambda\Omega^2\frac{B_4}{\eta^5}\,\bm{F}_4
\nonumber\\
&&\qquad
+\,\,2\lambda^{3/2}\Omega\frac{J_1}{\eta^2}\,\bm{Z}_1
+2\lambda^{3/2}\Omega^3\frac{J_3}{\eta^4}\,\bm{Z}_3\,,
\label{exg1}
\end{eqnarray}
where $\eta=r/r_0$ is a dimensionless variable. Let us remark that up to this order of approximation
$M_2$, $J_3$, $A_2$ and
$B_4$ are pure numbers but $M_0$, $J_1$ and $B_2$ can be 
linear functions of
$\Omega^2$.

\subsection{Matching  surface and  energy-momentum tensor}
Let us assume the interior metric has the same dependence on $\lambda$ 
than the exterior metric (\ref{exg1}) has,
\begin{eqnarray}
&&\gamma_{tt} \approx -1 + \lambda f_{tt}\,,\quad
\gamma_{t\varphi} \approx\lambda^{3/2}f_{t\varphi}\,,\quad
\gamma_{\varphi\varphi} \approx 1+\lambda f_{\varphi\varphi}\,,
\nonumber\\
&&\gamma_{rr} \approx 1 + \lambda f_{rr}\,,\quad 
\gamma_{r\theta} \approx\lambda f_{r\theta}\,,\quad
\gamma_{\theta\theta} \approx 1+\lambda f_{\theta\theta}
\end{eqnarray}
(we shall not use labels to distinguish between exterior or interior metrics not even between first or second order approximate metrics whenever it can be clearly understood to which of them we are referring).  Thus the normalization factor (\ref{eqnorma}) up to first order in $\lambda$ reads,
\begin{equation}
\psi \approx 1 + \frac{1}{2}\lambda \left(f_{tt} +\Omega^2 \eta^2\sin^2\theta\right)\,.
\label{psiapprox}
\end{equation}
As far we are assuming that the metric components are continuous 
on the matching surface, we use their exterior expressions  given by (\ref{exg1}) to make up (\ref{eqsuperficie}) into  a true equation for such a surface. Then
we search for a parametric form of the matching surface up to zeroth order in $\lambda$ by making the following assumption,
\begin{equation}
r\approx r_0\left(1+\sigma\Omega^2 P_2(\cos\theta)\right)\,.
\label{desenvradisup}
\end{equation}
A simple calculation leads to
\begin{equation}
\sigma\approx\frac{1}{M_0}\left(M_2-\frac13\right).
\label{eta21er}
\end{equation}
We also obtain a similar expression for  $\psi_\Sigma$ in terms of the exterior constants,
\begin{equation}
\psi_\Sigma\approx 1+\lambda\left(M_0 +\frac{1}{3}\Omega^2\right)\,.
\label{psis1}
\end{equation}

To complete the linear interior equations we need an approximate 
expression of the energy-momentum tensor of the fluid. First of all, let us notice that  the density
$\mu=3\lambda/(4\pi r_0^2)$ is a quantity of order 
$\lambda$. Thus taking into account (\ref{psiapprox}) and (\ref{psis1}), it is easy to check that the pressure (\ref{eqpressio1}) is of order
$\lambda^2$. Therefore the energy-momentum tensor (\ref{eqenermom}) contributes to the right side of the Einstein equations by means of
\begin{equation}
8\pi\,\bm{t} \approx
3\frac{\lambda}{ r_0^2}\left(\bm{T}_0+\bm{D}_0\right)
+6\frac{\lambda^{3/2}\Omega}{r_0^2}\eta\,\bm{Z}_1\,,
\label{impulsenerg0}
\end{equation}
if terms of order equal or higher than $\lambda^2$ are disregarded.

\subsection{Linear interior solution}
Let us consider the following system of linear differential equations
\begin{eqnarray}
&&\triangle h_{\alpha\beta} = -16\pi t_{\alpha\beta}\,,
\nonumber\\
[.6ex]
&&\partial^k (h_{k\mu} -\frac12 h\,\eta_{k\mu}\,) = 0\,,
\label{eqint}
\end{eqnarray}
where $\bm{t}$ is given by (\ref{impulsenerg0}). The general 
solution of these equations which is regular at the origin of the coordinate system $r=0$ and has the Papapetrou structure (\ref{eqmetrica}) is
\begin{eqnarray}
&&\bm{h} =\bm{h}_{\rm inh}+ \sum_{n=0}^\infty m_n\, r^n\left(\bm{T}_n+\bm{D}_n\right)
+ \sum_{n=1}^\infty \,j_n\, r^n\,\bm{Z}_n
\nonumber\\
&&\ \ \,
+\,\,a_0\,\bm{D}_0 +b_0\,\bm{H}_0
+\,\sum_{n=1}^\infty a_n r^n\bm{E^*}_n
+\sum_{n=2}^\infty b_n\, r^n\,\bm{F^*}_n \,,
\label{lininthar}
\end{eqnarray}
where
\begin{eqnarray}
\bm{h}_{\rm inh}=-\lambda\eta^2\,(\bm{T}_0+\bm{D}_0)-\frac65\lambda^{3/2}\Omega\eta^3\,\bm{Z}_1
\end{eqnarray}
is a solution of the inhomogeneous system. We have also 
introduced two new sets of spherical harmonic tensors,
\begin{eqnarray}
&& \bm{E^*}_n \equiv \frac23(n+1)\,\bm{D}_n-\frac23(n+1)\,\bm{H}_n +\bm{H}^1_n \quad (n\geq 1)\,,
\nonumber\\
&& \bm{F^*}_n \equiv \frac12(n+1)(n+2)\,\bm{H}_n-(n+2)\,\bm{H}^1_n -\frac12\,\bm{H}^2_n
\quad (n\geq 2)\,,
\end{eqnarray}
which seem to be suited to the interior problem better than those we used to write the linear 
exterior metric.

The integration constants that appear in expression (\ref{lininthar}) should have some dependence on the physical parameters $\lambda$, $\Omega$ and $r_0$ similar to that we
propose for the constants of the exterior metric (there we used capital letters to label the constants, here we use small case letters). We assign to $m_n$, $j_n$
and $a_n$ the same dependence on $\lambda$ and $\Omega$ that we assumed for their corresponding capitals. This rule must be slightly modified for the constants 
$b_n$, in the sense that they differ from their capitals in a factor $\Omega^2$ ($\bm{F^*}_2$ is not a spherical symmetric tensor). There are
two exceptions, $a_0$ and $b_0$, because they have not counterpart in the exterior metric, we assume that both are linear in $\lambda$ and zeroth order in
$\Omega$ (in fact $b_0$ must be of second order in $\Omega$ since $\bm{H}_0$ has not spherical symmetry).
Finally, we must include a factor
$r_0^{-n}$ in all constants with an index
$n$ as they are associated to
$r^n$ terms of the interior metric. It can be summarized as follows:
\begin{eqnarray}
&&m_0\rightarrow \lambda m_0\,,\quad
 m_2\rightarrow\lambda\Omega^2\frac{m_2}{r_0^2}\,,\quad
 j_1\rightarrow\lambda^{3/2}\Omega\frac{j_1}{r_0}\,,\quad
 j_3\rightarrow\lambda^{3/2}\Omega^3\frac{j_3}{r_0^3}\,,\quad
\nonumber\\
&&a_0 \rightarrow \lambda a_0\,,\quad
 b_0 \rightarrow\lambda b_0\,,\quad
 a_2 \rightarrow\lambda\Omega^2\frac{a_2}{r_0^2}\,,\quad
 b_2  \rightarrow \lambda\Omega^2\frac{b_2}{r_0^2}\,.
\label{constint}
\end{eqnarray}
Constants we have not mentioned above whether depend on higher powers of $\Omega$ or are incompatible with the equatorial symmetry requirement, so they 
can be ignored. As before, these new dimensionless constants are pure numbers, $m_2$, $j_3$, $a_2$ and $b_2$, or linear functions of $\Omega^2$, $m_0$, $j_1$, $a_0$ and $b_0$.

Finally, substituting these values for the constants into equation (\ref{lininthar}),
we arrive to the following approximate expression for the  interior
metric:
\begin{eqnarray}
&&\bm{g}_{\rm int}\approx
(-1+\lambda m_0-\lambda\eta^2)\,\bm{T}_0
+(1+\lambda m_0+\lambda a_0-\lambda\eta^2)\,\bm{D}_0
+\lambda b_{0}\,\bm{H}_0
\nonumber\\
&&\ \ \quad
+\,\,\lambda\Omega^2 m_2\eta^2\left(\bm{T}_2+\bm{D}_2\right)
+\lambda\Omega^2 a_2\eta^2\,\bm{E^*}_2
+\lambda\Omega^2 b_2\eta^2\,\bm{F^*}_2
\nonumber\\
&&\ \ \quad
+\,\,\lambda^{3/2}\Omega\eta\left( j_1-\frac65\eta^2\right)\bm{Z}_1
+\lambda^{3/2}\Omega^3 j_3\eta^3\,\bm{Z}_3\,.
\label{intg1}
\end{eqnarray}

\subsection{First order global solution}

Let us remember the matching conditions we use in this paper:  the metric components and their first derivatives have to be continuous through the  hypersurface
of zero pressure. Imposing these conditions on metrics (\ref{exg1}) and (\ref{intg1}) on the
surface given by (\ref {desenvradisup}) and (\ref{eta21er}) and keeping in mind the order of
approximation we are concerned, an straightforward calculation leads to the following values for
the exterior constants,
\begin{eqnarray}
&& M_0 \approx 1\,,\quad M_2\approx -\frac12\,,\quad
J_1 \approx  \frac25+\frac{1}{3}\Omega^2\,,\quad
J_3 \approx-\frac17\,,
\nonumber \\
&& A_2 \approx  B_2 \approx B_4 \approx 0\,,
\label{constant1}
\end{eqnarray}
these values for the interior constants,
\begin{eqnarray}
&& m_0 \approx 3\,,\quad m_2 \approx -1\,,\quad
j_1 \approx 2+\frac23\Omega^2\,,\quad
j_3 \approx - \frac27\,,
\nonumber \\
&& a_0 \approx b_0\approx a_2 \approx b_2\approx 0
\end{eqnarray}
and this expression,  $\sigma \approx -5/6$, for the unknown function in the surface equation. They all ensure that the matching has been properly done.

The above results lead to the following simple expression for the global metric:
\begin{eqnarray}
&&\bm{g}_{\rm int}\approx
(-1+3\lambda-\lambda\eta^2)\,\bm{T}_0+(1+3\lambda-\lambda\eta^2)\,\bm{D}_0
-\lambda\Omega^2\eta^2\left(\bm{T}_2+\bm{D}_2\right)
\nonumber \\
&&\qquad
+\,\,2\lambda^{3/2}\Omega\eta\left(1+\frac{1}{3}\Omega^2-\frac35\eta^2\right)\bm{Z}_1
-\frac27\lambda^{3/2}\Omega^3\eta^3\,\bm{Z}_3\,,
\nonumber\\
[.6ex]
&&\bm{g}_{\rm ext}\approx
\left(-1+2\frac{\lambda}{\eta}\right)\bm{T}_0
+\left(1+2\frac{\lambda}{\eta}\right)\bm{D}_0
-\frac{\lambda\Omega^2}{\eta^3}\left(\bm{T}_2+\bm{D}_2\right)
\nonumber\\
&&\qquad
+\,\,2\frac{\lambda^{3/2}\Omega}{\eta^2}\left(\frac25+\frac{1}{3}\Omega^2\right)\bm{Z}_1
-2\frac{\lambda^{3/2}\Omega^3}{7\eta^4}\,\bm{Z}_3\,.
\label{g1}
\end{eqnarray}

\section{Second order metric in harmonic coordinates}

\subsection{Non-linear exterior solution}

First of all we have to compute the right hand side of the equation (\ref{equiteracio}), that is
$\bm{H}$ and $\bm{S}$. We just need  approximate expressions of these quantities which
can be obtained by means of the first order metric (\ref{g1}) we worked out in the preceding
section. In a standard post-Minkowskian approximation, it deserves nothing
but a  direct calculation keeping in mind we have only to retain
terms of order less or equal to $\lambda^2$. However let us remember we have made
a slow rotation assumption, so we also have to drop out terms that depend on powers of
$\Omega$ higher than $\Omega^3$.  Thus we get
\begin{eqnarray}
&&\bm{S}\approx
 -4\frac{\lambda^2}{r_0^2\eta^4}\left(\bm{T}_0 -\frac23\,\bm{D}_0\right)
+4\frac{\lambda^2}{r_0^2\eta^4}\left(\frac13+\frac{\Omega^2}{7\eta^2}\right)\,\bm{F}_2
\nonumber\\
&&\qquad
 +\,\,4\frac{\lambda^2\Omega^2}{r_0^2\eta^6}\left(3\,\bm{T}_2-\frac{16}{7}\,\bm{D}_2
 -\frac{3}{14}\,\bm{E}_2-\frac17\,\bm{F}_4\right)\,,
 \nonumber\\
[.6ex]
&&\bm{H}\approx 4\frac{\lambda^2}{r_0\eta^3}\left(1+\frac{\Omega^2}{10\eta^2}\right)\bm{V}_1
-4\frac{\lambda^2}{r_0\eta^3}\left(1-\frac{\Omega^2}{5\eta^2}\right)\bm{W}_1
\nonumber\\
&&\qquad -\,\,12\frac{\lambda^2\Omega^2}{5r_0\eta^5}\left(\bm{V}_3-3\,\bm{W}_3\right)\,,
\end{eqnarray}
where
\begin{eqnarray}
&&\bm{V}_n=P_n^1(\cos\theta)\,k_idx^i \quad (n\geq 1)\,,
\nonumber\\
&&\bm{W}_n=P_n(\cos\theta)\,e_idx^i \quad (n\geq 0)
\end{eqnarray}
are spherical harmonic vectors.

Substituting these expressions of $\bm{S}$ and $\bm{H}$ into equations (\ref{equiteracio}), we have a
linear system of partial differential equations. It is a rather easy task to find a
solution of this problem  applying standard techniques, we choose that
\begin{eqnarray}
&&\bm{h}_{\rm inh} =
-2\frac{\lambda^2}{\eta^2}\left(\bm{T}_0-\frac23\,\bm{D}_0\right)
-\frac{\lambda^2}{3\eta^2} \left(1-2\frac{\Omega^2}{7\eta^2}\right)\bm{F}_2
\nonumber \\
&&\quad\quad
+\,\,\frac{\lambda^2\Omega^2}{\eta^4}\left(2\,\bm{T}_2-\frac{32}{21}\,\bm{D}_2-\frac17\,\bm{E}_2
+\frac{1}{14}\,\bm{F}_4\right)\,.
\label{solp2}
\end{eqnarray}
This is not the best choice we can make because we need a lot of arbitrary constants
if we attempt to match an exterior to an interior metric even in an approximate way.
The linear exterior solution (\ref{eqsolinexthar}), that is a
solution of the homogeneous system (\ref{equiteracio}), give us all the constants we
need. Therefore, our second order approximation to the exterior metric is a sum of
three terms: the first order global exterior metric (\ref{g1}) outside the source,
a part of the solution to the exterior linear problem, mainly 
(\ref{exg1}) after having dropped out the Minkowski metric and the $\bm{Z}_1$ and
$\bm{Z}_3$ terms and having multiplied the result by
$\lambda$ to obtain a second order expression, and the solution to the second order
post-Minkowskian system (\ref{solp2}). In fact, the first and second contributions to that expression can be obtained by substituting every constant in formula
(\ref{exg1}) by its value after first order matching (\ref{constant1}) plus a new constant times
$\lambda$ (we give to these new constant  the same name since they can be seen as corrections to the previous ones); that is,
\begin{eqnarray}
&&M_0 \rightarrow 1+\lambda M_0\,,\quad M_2 \rightarrow -\frac12+\lambda M_2\,,
\nonumber\\
&&A_2 \rightarrow \lambda A_2\,,\quad B_2 \rightarrow \lambda B_2\,,\quad B_4 \rightarrow \lambda B_4\,,
\label{susti2}
\end{eqnarray}
$J_1$ and $J_3$ have not to be corrected because we stop our calculation at order
$\lambda^2$. We arrive at the following expression for the exterior metric,
\begin{eqnarray}
&&\bm{g}_{\rm ext}\approx
\left(-1+2\frac{\lambda}{\eta}+2\lambda^2\frac{M_0}{\eta}\right)\bm{T}_0
+\left(1+2\frac{\lambda}{\eta}+2\lambda^2\frac{M_0}{\eta}\right)\bm{D}_0
\nonumber\\
&&\quad\quad
-\,\,\frac{\lambda\Omega^2}{\eta^3}\left(1-2\lambda M_2\right)\,\left(\bm{T}_2+\bm{D}_2\right)
+\lambda^2 \frac{B_2}{\eta^3}\,\bm{F}_2
\nonumber\\
&&\quad\quad
+\,\,\lambda^2\Omega^2 \frac{A_2}{\eta^3}\,\bm{E}_2
+\lambda^2\Omega^2 \frac{B_4}{\eta^5}\,\bm{F}_4
+\bm{h}_{\rm inh}
\nonumber\\
&&\quad\quad
+\,\,2\frac{\lambda^{3/2}\Omega}{\eta^2}\left(\frac25+\frac{1}{3}\Omega^2\right)\bm{Z}_1
-2\frac{\lambda^{3/2}\Omega^3}{7\eta^4}\,\bm{Z}_3\,.
\label{sol12}
\end{eqnarray}

As far as we have redefined the first order solution to introduce the
second order solution constants, we can say that there are not other
constants in our approach than those of the linear solution. Moreover,
they can be written as double series in powers of the source parameters
$\lambda$ and $\Omega$. This is the kind of expressions we shall find for the
multipole moments of the vacuum exterior metric.

\subsection{Non-linear interior solution}

First of all let us say something about the matching surface. The normalization factor (\ref{eqnorma}) up to order $\lambda^2$ reads
\begin{eqnarray}
&&\psi\approx
1+\frac{1}{2}\lambda\left(f_{tt}+\Omega^2\eta^2\sin^2\theta\right)
+\frac12\lambda^2\left[\frac34 f_{tt}^2 + k_{tt}\right.
\nonumber\\
&&\quad\ \ 
\left.+\,\,2\Omega f_{t\varphi}\,\eta\sin\theta
+\Omega^2\left(\frac32 
f_{tt}+f_{\varphi\varphi}\right)\eta^2\sin^2\theta\right],
\label{psi2order}
\end{eqnarray}
where $k_{tt}$ is the $\lambda^2$ term in the expansion of $\gamma_{tt}$. Following what we did at first order, we use the continuity of 
the metric components on the matching surface to obtain an equation
for this  surface from the above expression. From the formula (\ref{sol12}) we obtain the following expressions for the exterior metric components we need:
\begin{eqnarray}
&&  f_{tt}\approx \frac{2}{\eta}- \frac{\Omega^2}{\eta^3}P_2(\cos\theta)\,,
\nonumber\\
&&  k_{tt}\approx
\frac{2}{\eta}\left(M_0 - \frac{1}{\eta}\right)
+2\frac{\Omega^2}{\eta^3}\left(M_2+\frac{1}{\eta}\right)P_2(\cos\theta)\,,
\nonumber\\
&& f_{t\varphi}\approx 4\frac{\Omega}{5\eta^2}P_1^1(\cos\theta)\,,
\nonumber\\
&& f_{\varphi\varphi}\approx \frac{2}{\eta}-\frac{\Omega^2}{\eta^3}P_2(\cos\theta)\,.
\label{fab}
\end{eqnarray}
We substitute them into (\ref{psi2order}) and search for an explicit equation
of the matching surface as a slight perturbation of
the first order matching surface (\ref{desenvradisup}),
\begin{equation}
r\approx r_0\left[1+\left(-\frac56+\lambda\sigma\right)\Omega^2P_2(\cos\theta)\right]
\label{surf2}
\end{equation}
(the notation goes in the same way we have been using). Then, we find
\begin{equation}
\sigma \approx -\frac45+\frac56 M_0+M_2\,.
\label{sig2}
\end{equation}

The value of $\psi$ on the matching surface up to the order we need is
already known, we have just to set  $M_0=1$, its first order matching
value, in its expression (\ref{psis1}). Moreover $\psi$ is known too. We simply read out $f_{tt}$ form the expression of the first order global metric inside the
source (\ref{g1}) and substitute it into (\ref{psiapprox}) to obtain it. Therefore we also have the expression for the pressure we need to write down the
energy-momentum tensor up to order $\lambda^2$. That is,
\begin{equation}
p \approx \frac{\lambda^2}{8\pi r_0^2}\left((3-2\Omega^2)(1-\eta^2)- 5\Omega^2\eta^2 P_2(\cos\theta)\right)\,.
\label{presion}
\end{equation}

\bigskip

Let us  proceed to obtain the second order interior solution. First we evaluate the right hand side of equation (\ref{equiteracio})
using the known first order solution (\ref{g1}). Now this task implies
not only to obtain the non-linear terms in the metric but the
energy-momentum tensor up to second order, i.~e. neglecting terms in
$\lambda^{5/2}$ and $\Omega^4$. The result is the following:
\begin{eqnarray}
&&\bm{S}\approx
-\frac{\lambda^2}{r_0^2}\left(9-6\Omega^2-5\eta^2+14\Omega^2\eta^2\right)\,\bm{T}_0
\nonumber\\
&&\quad
 -\,\frac{\lambda^2}{r_0^2}\left(33+2\Omega^2-\frac{35}{3}\eta^2+\frac23\Omega^2\eta^2\right)\,\bm{D}_0
 +4\frac{\lambda^2}{3r_0^2}(1-4\Omega^2) \eta^2\,\bm{F}_2
\nonumber\\
&&\quad
+\,\,\frac{\lambda^2\Omega^2}{r_0^2}\eta^2\left(15\,\bm{T}_2+23\,\bm{D}_2+6\,\bm{E}_2\right)\,,
\nonumber\\
[.6ex]
&&\bm{H}\approx
 2\frac{\lambda^2}{r_0}\eta\left(3-\frac32\Omega^2-\eta^2+\frac{7}{10}\Omega^2\eta^2\right)\bm{V}_1
 -2\frac{\lambda^2}{r_0}\eta\left(3+3\Omega^2\right.
\nonumber\\
&&\qquad 
\left.-\,\,\eta^2-\frac75\Omega^2\eta^2\right)\bm{W}_1
-2\frac{\lambda^2\Omega^2}{5r_0}\eta^3\left(\bm{V}_3-3\,\bm{W}_3\right)
\label{SQ2}
\end{eqnarray}
Then, we build up the second order interior solution with the same ingredients
we did the exterior solution: we obtain an exact solution of the
inhomogeneous system defined by equations (\ref{equiteracio}) and (\ref{SQ2}),
\begin{eqnarray}
&&\bm{h}_{\rm inh} =
\lambda^2\eta^2\left(-\frac32 +\Omega^2+\frac{1}{4}\eta^2-\frac{7}{10}\Omega^2\eta^2\right)\bm{T}_0
\nonumber\\
&&\qquad
-\,\,\lambda^2\eta^2\left(\frac{11}{2}+\frac{1}{3}\Omega^2-\frac{7}{12}\eta^2+\frac{1}{30}\Omega^2\eta^2\right)\bm{D}_0
\nonumber\\
&&\qquad
+\,\,\lambda^2\eta^2\left(-\frac15+\frac43\Omega^2+\frac{2}{21}\eta^2-\frac{8}{21}\Omega^2\eta^2\right)\bm{F}_2
\nonumber\\
&&\qquad
+\,\,3\lambda^2\Omega^2\eta^2\left(- \frac35+\frac{1}{7}\eta^2\right)\bm{E}_2
\nonumber\\
&&\qquad
+\,\,\frac{1}{14}\lambda^2\Omega^2\eta^4\left(15\,\bm{T}_2+23\,\bm{D}_2+\frac{1}{9}\,\bm{F}_4\right)\,,
\end{eqnarray}
to which we add an adapted version of the general solution of the
homogeneous system
and the first order interior solution (\ref{g1}). Thus we get
\begin{eqnarray}
&&\bm{g}_{\rm int}\approx
(-1+3\lambda-\lambda\eta^2+\lambda^2m_0)\,\bm{T}_0
+(1+3\lambda-\lambda\eta^2+\lambda^2 m_0)\,\bm{D}_0
\nonumber\\
&&\qquad
-\,\,\lambda\Omega^2\left(1-\lambda m_2\right)\eta^2\left(\bm{T}_2+\bm{D}_2\right)
+\lambda^2 a_0\,\bm{D}_0
+\lambda^2 b_{0}\,\bm{H}_0
\nonumber\\
&&\qquad
+\,\,\lambda^2\Omega^2 a_2\eta^2\,\bm{E^*}_2
+\lambda^2\Omega^2 b_2\eta^2\,\bm{ F^*}_2
+\bm{h}_{\rm inh}
\nonumber\\
&&\qquad
+\,\,2\lambda^{3/2}\Omega\eta\left(1+\frac13\Omega^2-\frac35\eta^2\right)\bm{Z}_1
-\frac27\lambda^{3/2}\Omega^3\eta^3\,\bm{Z}_3\,.
\label{int2}
\end{eqnarray}
Here the free constants can also be thought as corrections of
order $\lambda$ to the the first order interior 
solution constants,  in the same way
we suggested when we were solving the non-linear exterior problem,
\begin{eqnarray}
&& m_0 \rightarrow 3+ \lambda m_0\,,\quad m_2 \rightarrow -1+\lambda m_2\,,
\nonumber\\
&&a_0 \rightarrow \lambda a_0\,,\quad b_0 \rightarrow \lambda b_0\,,\quad
a_2 \rightarrow \lambda a_2\,,\quad b_2 \rightarrow \lambda b_2\,,
\end{eqnarray}
$j_1$ and $j_3$ do not need to be changed.

\subsection{Second order global solution}

Second order interior metric (\ref{int2}) can be matched to the exterior metric (\ref{sol12}) on the surface defined by (\ref{surf2}) and (\ref{sig2})  if an only
if
\begin{eqnarray}
&& M_0\approx 3 +\frac25\Omega^2\,,\quad M_2 \approx -\frac{69}{70}\,,\nonumber\\
&& A_2 \approx 0\,,\quad B_2 \approx \frac8{35}\left(1 +\frac16\Omega^2\right)\,,\quad
 B_4 \approx -\frac4{63}
 \label{M2}
\end{eqnarray}
and 
\begin{eqnarray}
&&m_0 \approx \frac14(21+2\Omega^2)\,,\quad m_2 \approx -\frac{73}{70}\,,\nonumber\\
&& a_0 \approx \frac13(21+2\Omega^2)\,,\quad b_0 \approx 0\,,\quad
a_2 \approx -\frac{43}{15}\,,\quad b_2 \approx -\frac{86}{105}\,.
\end{eqnarray}
Substituting (\ref{M2}) into (\ref{sig2}), we obtain the definite equation of the matching surface,
\begin{equation}
r\approx r_0\left[1-\frac56\left(1-\frac67\lambda\right)\Omega^2P_2(\cos\theta)\right]\,.
\end{equation}

The global second order solution reads,
\begin{eqnarray}
&&\bm{g}_{\rm int}\approx
\left[-1+3\lambda-\lambda\eta^2+\frac{1}{4}\lambda^2\left(21 +2 \Omega^2\right)-\lambda^2\left(\frac32-\Omega^2\right)\eta^2\right.
\nonumber\\
&&\quad
\left.+\,\,\frac{1}{2}\lambda^2\left(\frac12-\frac{7}{5}\Omega^2\right)\eta^4\right]\bm{T}_0
+\left[1+3\lambda-\lambda\eta^2+\frac{7}{12}\lambda^2\left(21 +2\Omega^2\right)\right.
\nonumber\\
&&\quad
\left. -\,\,\lambda^2\left(\frac{11}2+\frac{1}{3}\Omega^2\right)\eta^2
+\frac{1}{6}\lambda^2\left(\frac72-\frac{1}{5}\Omega^2\right)\eta^4\right]\bm{D}_0
\nonumber\\
&&\quad
-\,\,\lambda\Omega^2\eta^2\left(1+\frac{73}{70}\lambda-\frac{15}{14}\lambda\eta^2
\right)\bm{T}_2
-\lambda\Omega^2\eta^2\left(1+\frac{1079}{210}\lambda-\frac{23}{14}\lambda\eta^2\right)\bm{D}_2
\nonumber\\
&&\quad
-\,\,\lambda^2\eta^2\left[\frac15\left(1-\frac{18}{7}\Omega^ 2\right) -\frac{2}{21}
\left(1- 4\Omega^2\right)\eta^2\right]\bm{F}_2
\nonumber\\
&&\quad
-\,\,\frac17\lambda^2\Omega^2\eta^2\left(4-3\eta^2\right)\bm{E}_2
+\frac{1}{126}\lambda^2\Omega^2\eta^4\,\bm{F}_4
\nonumber\\
&&\quad
+\,\,2\lambda^{3/2}\Omega\eta\left(1+\frac13\Omega^2-\frac35\eta^2\right)\bm{Z}_1-\frac27\lambda^{3/2}\Omega^3\eta^3\,\bm{Z}_3\,,
\nonumber\\ 
[.6ex]
&&\bm{g}_{\rm ext}\approx
\left[-1+2\frac{\lambda}{\eta}+2\frac{\lambda^2}{\eta}\left(3
+\frac25\Omega^2\right)-2\frac{\lambda^2}{\eta^2}\right]\bm{T}_0\nonumber\\
&&\qquad
 +\,\left[1+2\frac{\lambda}{\eta}+2\frac{\lambda^2}{\eta}\left(3
+\frac25\Omega^2\right)+4\frac{\lambda^2}{3\eta^2}\right]\bm{D}_0
\nonumber\\
&&\qquad
-\,\,\frac{\lambda\Omega^2}{\eta^3}\left(1+\frac{69}{35}\lambda-2\frac{\lambda}{\eta}\right)\bm{T}_2
-\frac{\lambda\Omega^2}{\eta^3}\left(1+\frac{69}{35}\lambda+32\frac{\lambda}{21\eta}\right)\bm{D}_2
\nonumber\\
&&\qquad 
-\,\,\frac{\lambda^2}{\eta^2}\left[\frac13-\frac8{35\eta}\left(1
+\frac16\Omega^2\right)-2\frac{\Omega^2}{21\eta^2}\right]\bm{F}_2
\nonumber\\
&&\qquad
-\,\,\frac{\lambda^2\Omega^2}{7\eta^4}\,\bm{E}_2
+\frac{\lambda^2\Omega^2}{14\eta^4}\left(1-\frac{8}{9\eta}\right)\bm{F}_4
\nonumber\\
&&\qquad
+\,\,2\frac{\lambda^{3/2}\Omega}{\eta^2}\left(\frac25+\frac13\Omega^2\right)\bm{Z}_1
-2\frac{\lambda^{3/2}\Omega^3}{7\eta^4}\,\bm{Z}_3\,.
\label{g2}
\end{eqnarray}

Let us remark that the first order metric plays a relevant role in the
above matching. When we evaluate the first order metric or their
derivatives on the matching surface, substituting $r$ by its
expression (\ref{surf2}), there are generated terms of order
higher than
$\lambda$ that can not be ignored in the matching process.

To end this subsection let us say something about the constants we have
used to make the Lichnerowicz matching. The more interesting among
them are obviously $M_0$, $M_2$, $J_1$ and $J_3$. They are
Thorne-Geroch-Hansen multipole moments, so they form part of an
intrinsic (coordinate independent) characterization of vacuum
asymptotically flat space-times. We must go back to the very beginning of
the paper and try to join all the pieces into which we have split those
constants if we want to write down them as they were introduced in formula
(\ref{eqsolinexthar}). The tour starts in equation (\ref{substitution}), goes
through formulas (\ref{constant1}), (\ref{susti2}), and (\ref{M2}) to arrive at
\begin{eqnarray}
&&M_0 \approx \lambda r_0\left[1 + \lambda\left(3
+\frac25\Omega^2\right)\right]\,,\nonumber\\
&&M_2 \approx -\frac12\lambda\Omega^2r_0^3\left(1 +\frac{69}{35} 
\lambda\right)\,,\nonumber\\
&&\ \,
J_1 \approx \frac25\lambda^{3/2}\Omega r_0^2\left(1+\frac56\Omega^2\right)\,,\nonumber\\
&&\ \,
J_3 \approx -\frac17\lambda^{3/2}\Omega^3r_0^4\,,
\label{moments}
\end{eqnarray}
and
\begin{eqnarray}
A_2\approx 0\,,\quad
B_2\approx\frac{8}{35}\lambda^2r_0^3\left(1+\frac16\Omega^2\right)\,,\quad
B_4\approx -\frac{4}{63}\lambda^2\Omega^2r_0^5\,.
\label{tensiones}
\end{eqnarray}
Even though $A_2$, $B_2$ and $B_4$ are not intrinsic, they may
be of some interest since they are set up in an unambiguous way
by the Lichnerowicz matching conditions: they may carry some extra information about the
source of the gravitational field, that is to say the kind of matter we are
considering.

\subsection{Schwarzschild and Kerr metrics}

The static limit of our metric (\ref{g2}),
\begin{eqnarray}
&&\bm{g}_{\rm int}\approx
\left(-1+3\lambda-\lambda\eta^2+\frac{21}{4}\lambda^2-\frac32\lambda^2\eta^2
+\frac14\lambda^2\eta^4\right)\bm{T}_0
\nonumber\\
&&\qquad
+\,\,\left(1+3\lambda-\lambda\eta^2+\frac{49}{4}\lambda^2-\frac{11}{2}\lambda^2\eta^2
+\frac{7}{12}\lambda^2\eta^4\right)\bm{D}_0
\nonumber\\
&&\qquad
-\,\,\lambda^2\eta^2\left(\frac15 -\frac{2}{21}\eta^2\right)\bm{F}_2\,,
\nonumber\\
[.6ex]
&&\bm{g}_{\rm ext}\approx
\left(-1+2\frac{\lambda}{\eta}+6\frac{\lambda^2}{\eta}-2\frac{\lambda^2}{\eta^2}\right)\bm{T}_0
+\left(1+2\frac{\lambda}{\eta}+6\frac{\lambda^2}{\eta}+4\frac{\lambda^2}{3\eta^2}\right)\bm{D}_0
\nonumber\\
&&\qquad 
-\,\,\frac{\lambda^2}{\eta^2}\left(\frac13-\frac8{35\eta}\right)\bm{F}_2\,,
\label{Schwarz}
\end{eqnarray}
must be an approximation to the global metric that results from matching Schwarzschild's metric
to its well known interior metric (an exact solution of the Einstein
equations for a perfect fluid having constant mass
density) \cite{Malcolm}. However, this metric is usually written in standard coordinates, therefore a direct comparison of this metric to (\ref{Schwarz}) is not allowed. Nevertheless, it has
recently been shown \cite{Jose Luis} that Schwarzschild's metric
admits a global system of harmonic coordinates. Taking into account some formulas given in \cite{Jose Luis}, we can work out an approximate change
of the radial coordinate which brings the Schwarzschild metric in standard coordinates to the form shown in (\ref{Schwarz}). It reads,
\begin{equation}
R(\eta)\approx \left\{
\begin{array}{ll}\displaystyle
r_0\eta\left[1+\frac{\lambda}{2}\left(3-\eta^2\right)+\lambda^2\left(5-\frac{21}{10}\eta^2+\frac{3}{14}\eta^4\right)\right],
& \eta\leq 1\\
&\\
\displaystyle
r_0\eta\left[1+\frac{\lambda}{\eta}+\frac{\lambda^2}{\eta}\left(3+\frac{4}{35\eta^2}\right)\right],
& \eta > 1\\
\end{array}
\right.
\label{standard}
\end{equation}
where $R$ is the standard radial coordinate.

Equation (\ref{standard}) also relates the source radius in standard coordinates to the source radius in harmonic coordinates,
that is,
\begin{equation}
R_0\equiv R(1)\approx r_0\left(1+\lambda+\frac{109}{35}\lambda^2\right)\,.
\end{equation}
We need this information to rewrite the Schwarzschild metric in harmonic coordinates, even outside the source, since the Schwarzschild mass is a function of $R_0$,
\begin{equation}
M=\frac43\pi\mu R_0^3\approx \lambda r_0(1+3\lambda)\,.
\end{equation}
$M$ is equal to the mass monopole moment $M_0$ as expected.

We can establish that Kerr's metric does not fit our exterior metric by comparing the first multipole
moments of the former \cite{Thorne1, ABMMR},
\begin{equation}
M_0=m\,,\quad M_2=-ma^2\,,\quad J_1=ma\,,\quad J_3=-\frac13 ma^3\,,
\label{Kerr}
\end{equation}
(here $m$ and $a$ stand for the usual parameters of the Kerr metric)
to those of our metric. They cannot be matched for
any choice of the free parameters $\lambda$, $\Omega$ and $r_0$: the following
combination of multipole moments,
\begin{equation}
\frac{J_1}{M_0}-3\frac{J_3}{M_2}\approx -\frac{16}{35}\lambda^{1/2}\Omega r_0
\end{equation}
does not vanish in our approach but it is equal to zero if we
evaluate it using the Kerr multipole moments.

\subsection{McLaurin ellipsoids}

The linear equation (\ref{eqint}) that must be satisfied by the component of the metric
deviation $h_{tt}$ looks like the equation for the Newtonian
gravitational field of a compact body with constant mass density
$\mu$. The classical solution to this problem is given by Poisson's
integral; that is, a scalar field which tends to zero at infinity and
such that the field and its first derivatives are continuous on the
surface of the body. Moreover, we know that the field outside of the body
is uniquely determined by some constants, the multipole
moments of the mass distribution.

Let us consider an
axisymmetric ellipsoid whose center is at the origin of the coordinate
system and its axis lies on the $z$-axis. Then its
mass and quadrupole moment are
\begin{eqnarray}
&& M=\frac{4\pi}{3}\mu a^2 b\,,\nonumber\\
&& Q=\frac15 M (b^2-a^2)\,,
\end{eqnarray}
where $a$ and $b$ stand, respectively, for the length of  the ellipsoid
semi-axis on the $xy$-plane and on the $z$-axis. These two parameters
are constrained in a self-gravitating rotating ellipsoid by the equation
(see for instance \cite{Landau})
\begin{equation}
\frac{\delta(1+2\delta^2)}{(1-\delta^2)^{3/2}}\arccos\delta
-\frac{3\delta^2}{1-\delta^2}-\frac23\Omega^2=0\,,
\end{equation}
where $\delta\equiv b/a$, and $\Omega$ is related to the constant angular
velocity of rotation of the ellipsoid $\omega$ in the same way as the
control  parameter we have been using. Solving that equation for
small values of
$\Omega$,
\begin{equation}
\delta \approx 1-\frac54\Omega^2\,,
\end{equation}
and introducing for convenience a new parameter $r_0$ defined by
$a=r_0\delta^{-1/3}$, we obtain
\begin{eqnarray}
&& M=\frac{4\pi}{3}\mu r_0^3=\lambda r_0,,\nonumber\\
&& Q=\frac15 \lambda r_0^3\delta^{-2/3}(\delta^2-1)
\approx -\frac12\lambda\Omega^2 r_0^3\,.
\end{eqnarray}
If we identify the parameters $\mu$, $\omega$ and $r_0$ of the
ellipsoid with those of our approximation scheme, these two expressions become identical  with the $\lambda$ terms of the multipole
moments $M_0$ and $M_2$ corresponding to our approximate
metric. Therefore, we can say
that the Newtonian limit of the metric (\ref{g2}) describes the
gravitational field of a slowly rotating McLaurin's ellipsoid.

Let us point out that the first dynamical multipole moment $J_1$ is just
the McLaurin ellipsoid angular momentum,
\begin{equation}
J=\frac25 Ma^2\omega= \frac25\lambda^{3/2}\Omega r_0^2\delta^{-2/3}
\approx \frac25\lambda^{3/2}\Omega r_0^2\left(1+\frac56\Omega^2\right)\,,
\end{equation}
and $J_3$ can also be related to the ``quadrupole moment'' of the
angular momentum density of the McLaurin ellipsoid.

\section{Quo-harmonic coordinates}

Quo-harmonic coordinates were introduced some years ago by Ll. Bel (see
\cite{ABMMR} for a brief review). They share with harmonic coordinates the
property of being an admissible set of local coordinates for any
space-time, since they are both defined by differential constraints that
have sense whatever the metric is. However, quo-harmonic coordinates
unlike harmonic coordinates require a timelike vector field to be
properly defined. That field is used to set the time coordinate and to
introduce the projector into the plane orthogonal to the field (quotient
metric), which in turn provides spatial
coordinates  as three independent harmonic functions of that three
dimensional positive definite Riemannian metric (quo-harmonic
coordinates). If these functions are chosen appropritely, they may look
like Cartesian coordinates in the same grounds as harmonic
coordinates can be thought of as a generalization of  Minkowski's coordinates
to curved space-times.

It is sometimes possible to
choose these coordinates as first integrals of the timelike vector
field. That property singles out a class of vector fields in any
Riemannian manifold \cite{Coll}. Those fields yield to introduce what
Bel calls quo-harmonic congruences, an attempt to understand rigid
motion in a framework less restrictive than that of Born's \cite{Llosa1, Llosa2}.

\subsection{Quo-harmonic coordinates and post-Minkowskian approximation }

The timelike Killing vector $\bm{\xi}$ defines a quo-harmonic congruence
in any space-time whose metric is given by (\ref{eqmetrica}). The
assumptions we have made on the coordinates (see section $2$) identify the
$t$ coordinate with the time we can associate with that Killing
field, since $\bm{\xi}$ has canonical form in this system of
coordinates. This coordinate is an harmonic function of the four dimensional metric. This constraint on the
time coordinate is not required by the quo-harmonic scenario, which
does not define this coordinate, but it is given as a bonus by the Papapetrou structure we
assume for the metric. Moreover, the projector operator associated with
$\bm{\xi}$
\begin{equation}
\bm{g}_{\xi}=\bm{g}-\bm{g}\left(\bm{\xi},\bm{\xi}\right)^{-1}\bm{\xi}\otimes\bm{\xi}\,,
\end{equation}
does not depend on $t$: it is a true three-dimensional metric. This is
more evident in our notation,
\begin{eqnarray}
&&\bm{g}_{\xi}=\gamma_{rr}\,\bm{\omega}^r{\otimes\,}\bm{\omega}^r+
\gamma_{r\theta}(\bm{\omega}^r{\otimes\,}\bm{\omega}^\theta+\bm{\omega}^\theta{\otimes\,}\bm{\omega}^r) 
+\gamma_{\theta\theta}\,\bm{\omega}^\theta{\otimes\,}\bm{\omega}^\theta
\nonumber\\
&&\qquad
+\,\,\gamma_{tt}^{-1}
\left(\gamma_{tt}\gamma_{\varphi\varphi}-\gamma_{t\varphi^2}\right)\,
\bm{\omega}^\varphi{\otimes\,}\bm{\omega}^\varphi
\equiv {g_{\xi}}_{ij}\,dx^i{\otimes\,}dx^j\,,
\label{qmetric}
\end{eqnarray}
where $\{x^i\}$ are the Cartesian-like coordinates we introduce in
section $2$. Then, if we want them to be quo-harmonic coordinates, we
have to impose the following condition on the Riemannian connection of the three dimensional metric
(\ref{qmetric})
\begin{equation}
{g_{\xi}}^{jk}{\Gamma_{\xi}}_{jk}^i=0
\end{equation}

From the point of view of a post-Minkowskian perturbation
approach, this coordinate condition is as admissible as the harmonic
condition (\ref{harmoniceq}). We obtain a well defined system of differential
equations that can be formally solved by iteration. The first-order approximation to the metric unambiguously determines the
inhomogeneous part of the system of equations which can be solved to work
out the second-order metric and so on.

The starting point is a system of linear partial differential
equations that are slightly more complex than those mentioned before, say (\ref{eqhomog}) or
(\ref{eqint}),
\begin{eqnarray}
&&\triangle h_{00} = -16\pi t_{00}\,,
\nonumber\\
[.6ex]
&&\triangle h_{0i} = -16\pi t_{0i}\,,
\nonumber\\
[.6ex]
&&\triangle h_{ij}-\partial_{ij}h_{00}  = -16\pi t_{ij}\,,
\nonumber\\
[.6ex]
&&\partial^k\left(h_{ki}-\frac12\delta_{ki}\delta^{lm}h_{lm}\right)=0\,.
\end{eqnarray}
Their solution can be written using the basis of
spherical harmonic tensors we introduced in the harmonic case.
Outside the source, $t_{\alpha\beta}=0$, it reads,
\begin{eqnarray}
&&\bm{h} =
2\sum_{n=0}^\infty\frac{M_n}{r^{n+1}}
\left(\bm{T}_n+2\frac{n+2}{2n+3}\,\bm{D}_n+\frac{1}{2(2n+3)}\,\bm{F}_{n+2}\right)
\nonumber\\
&&\quad
+\,\,2\sum_{n=1}^\infty\frac{J_n}{r^{n+1}}\,\bm{Z}_n
+\sum_{n=1}^\infty\frac{A_n}{r^{n+1}}\,\bm{E}_n
+\sum_{n=2}^\infty\frac{B_n}{r^{n+1}}\,\bm{F}_n\,;
\label{quolin1}
\end{eqnarray}
inside the source, it reads
\begin{eqnarray}
&&\bm{h} =\bm{h}_{\rm inh}
+m_0\,\bm{T}_0+m_1r\,\bm{T}_1+m_2r^2\left(\bm{T}_2+\frac23\,\bm{D}_2-\frac16\,\bm{H}_0\right)
\nonumber\\
&&\quad
+\,\,m_3r^3\left(\bm{T}_3+\frac45\,\bm{D}_3-\frac{3}{10}\left(\bm{H}_1-\bm{H}_1^1\right)\right)
\nonumber\\
&&\quad
+\,\sum_{n=4}^\infty m_nr^n
\left(\bm{T}_n+2\frac{n-1}{2n-1}\,\bm{D}_{n}
-\frac{1}{2(2n-1)}\,\bm{F}^{*}_{n-2}\right)
+\sum_{n=1}^\infty j_nr^n\,\bm{Z}_n
\nonumber\\
&&\quad
+\,\,a_0\,\bm{D}_0 +b_0\,\bm{H}_0
+\sum_{n=1}^\infty a_n r^n\,\bm{E}^{*}_n
+\sum_{n=2}^\infty b_nr^n\,\bm{F}^{*}_n\,.
\label{quolin2}
\end{eqnarray}
The notation for the constants is the same as that used  in the harmonic
case because they are required to have the same structure on $\lambda$ and $\Omega$ we
assumed there:  constants that are not compatible with equatorial
symmetry must be set equal to zero, the other constants are functions of the
physical parameters, $\lambda$, $\Omega$, and $r_0$ like those we
introduced by formulas (\ref{substitution}) and (\ref{constint}). This implies
that $\bm{t}$ is given by (\ref{impulsenerg0}) up to the first order in
$\lambda$ (even to order $\lambda^{3/2}$). Though its formal expression coincides with that of the harmonic case, of course it does not lead to the same
inhomogeneous solution,
\begin{equation}
\bm{h}_{\rm inh} =
-\lambda\eta^2\left(\bm{T}_0+\frac43\,\bm{D}_0-\frac{2}{15}\,\bm{F}_2\right)
-\frac65\lambda^{3/2}\Omega\eta^3\,\bm{Z}_1\,.
\end{equation}

Following the scheme we introduced in the harmonic case, we obtain the first-order
exterior and interior metrics by substituting (\ref{substitution}) and (\ref{constint}) into expressions (\ref{quolin1}) and (\ref{quolin2}) and neglecting terms
in $\lambda^2$ and $\Omega^4$, we find out the equation of the matching surface at this order of approximation, we impose the continuity of the metric
on that surface, etc\dots The  procedure is as before.
Therefore, we shall merely report the final results and comment on them.

\subsection{Second-order global solution}

The metric can be written  in quo-harmonic coordinates as follows:
\begin{eqnarray}
&&\bm{g}_{\rm int} \approx
 \left[-1+3\lambda-\lambda\eta^2+\lambda^2\left(\frac{153}{20}+\frac12\Omega^2\right)-\lambda^2\left(\frac52-\Omega^2\right)\eta^2
 \right.
 \nonumber\\
 &&\qquad
 \left. +\,\,\frac{\lambda^2}{10}\left(\frac92-7\Omega^2\right)\eta^4\right]\bm{T}_0
 +\left[1+4\lambda-\frac43\lambda\eta^2+\lambda\left(\frac{223}{10}+\frac43\Omega^2\right)
 \right.
 \nonumber\\
 &&\qquad
 \left.
 -\,\,\frac{32}{3}\lambda^2\eta^2+\frac{4}{15}\left(\frac{28}{5}-\Omega^2\right)\eta^4\right] \bm{D}_0
 +\frac{2}{15}\lambda\eta^2\left[1-\frac{5}{7}\Omega^2\right.
\nonumber\\
&&\qquad
\left.
+\,\,2\lambda\left(4-\frac{41}{21}\Omega^2\right)-\frac{\lambda}{7}\left(\frac{68}{5}-\frac{131}{14}\Omega^2\right)\eta^2
\right]\bm{F}_2
\nonumber\\
&&\qquad
-\,\,\frac16\lambda\Omega^2\left(1-\eta^2+\frac{521}{70}\lambda-\frac{2207}{210}\lambda\eta^2
+\frac{331}{100}\lambda\eta^4\right)\bm{H}_0
\nonumber\\
&&\qquad
- \,\,\lambda\Omega^2 \eta^2\left(1+\frac{499}{210}\lambda-\frac{109}{70}\lambda\eta^2\right)\bm{T}_2
\nonumber\\
&&\qquad
-\,\,\frac87\lambda\Omega^2\eta^{2}\left(1+\frac{185}{24}\lambda-\frac{18}{7}\lambda\eta^2\right)\bm{D}_{2}
\nonumber\\
&&\qquad
+\,\,\frac17\lambda\Omega^2 \eta^2\left(1+\frac{82}{15}\lambda-\frac{103}{70}\lambda\eta^2\right)\bm{E}_2
-\frac{169}{7350}\lambda^2\Omega^2\eta^4\,\bm{F}_4
\nonumber\\
&&\qquad
 +\,\,2\lambda^{3/2}\Omega\eta\left(1+\frac13\Omega^2-\frac35\eta^2\right)\bm{Z}_1
 -\frac27 \lambda^{3/2}\Omega^3 \eta^3\,\bm{Z}_3\,,
\nonumber\\
[.6 ex]
&&\bm{g}_{\rm ext} \approx
\left[-1+2\frac{\lambda}{\eta}+2\frac{\lambda^2}{5\eta}\left(21+2\Omega^2\right)
-3\frac{\lambda^2}{\eta^2}+\frac{\lambda^2}{5\eta^4}\right]\bm{T}_0
\nonumber\\
&&\quad
+\,\left[1+8\frac{\lambda}{3\eta}+8\frac{\lambda^2}{5\eta}\left(7+\frac23\Omega^2\right)
+11\frac{\lambda^2}{6\eta^2}+\frac{\lambda^2}{30\eta^4}
+3\frac{\lambda^2}{50\eta^6}\right]\bm{D}_0
\nonumber\\
&&\quad
+\,\,\frac{\lambda}{\eta}\left[
\frac{1}{3}-\frac{1}{\eta^2}\left(\frac15+\frac{2}{21}\Omega^2\right)
+\frac{\lambda}{5}\left(7+\frac23\Omega^2\right)+5\frac{\lambda}{12\eta}\right.
\nonumber\\
&&\quad
\left.
-\,\,\frac{\lambda}{35\eta^2}\left(31+\frac{102}{7}\Omega^2\right)
-\frac{\lambda}{3\eta^3}\left(\frac{1}{10}+\frac{4}{21}\Omega^2\right)
-\frac{\lambda}{5\eta^5}\left(\frac{9}{20}-\frac{16}{49}\Omega^2\right)
\right.
\nonumber\\
&&\quad
\left.
-\,\,3\frac{\lambda\Omega^2}{49\eta^7}\right]\bm{F}_2
-\,\,\frac{\lambda\Omega^2}{\eta^3}\left(1+\frac{139}{35}\lambda
-11\frac{\lambda}{3\eta}+18\frac{\lambda}{35\eta^3}\right)\bm{T}_2
\nonumber\\
&&\quad
-\,\,\frac{\lambda\Omega^2}{7\eta^3}\left(8+\frac{1112}{35}\lambda
+23\frac{\lambda}{3\eta}+5\frac{\lambda}{7\eta^3}
+33\frac{\lambda}{35\eta^5}\right)\bm{D}_2
\nonumber\\
&&\quad
+\,\,\frac{\lambda\Omega^2}{7\eta^3}\left(1+\frac{118}{35}\lambda
+2\frac{\lambda}{3\eta}-24\frac{\lambda}{35\eta^3}
+9\frac{\lambda}{14\eta^5}\right)\bm{E}_2
\nonumber\\
&&\quad
-\,\,\frac{\lambda\Omega^2}{14\eta^3}\left(
1-\frac{1}{\eta^2}
+\frac{139}{35}\lambda+9\frac{\lambda}{10\eta}
-67\frac{\lambda}{15\eta^2}+118\frac{\lambda}{175\eta^3}
-53\frac{\lambda}{70\eta^5}\right)\bm{F}_4
\nonumber\\
&&\quad
-\,\,\frac{\lambda^2\Omega^2}{20\eta^4}\left(\frac{29}{18}-\frac{13}{35\eta^2}
-\frac{3}{7\eta^4}\right)\bm{H}_0
\nonumber\\
&&\quad
+\,\,2\frac{\lambda^{3/2}\Omega}{\eta^2}\left(\frac25+\frac13\Omega^2\right)\bm{Z}_1
-2\frac{\lambda^{3/2}\Omega^3}{7\eta^4}\bm{Z}_3\,.
\label{quo2}
\end{eqnarray}

The equation of the matching surface, on which the metric and their first derivatives are continuous up to the order of approximation we are concerned, is
\begin{equation}
r\approx r_0\left[1-\frac56\left(1-\frac{22}{35}\lambda\right)\Omega^2P_2(\cos\theta)\right]\,.
\end{equation}
The pressure of the fluid has the same expression we found in the
harmonic case but obviously the meaning of the
coordinates and the parameters is not the same. It suggests that
quo-harmonic coordinates differ from  harmonic coordinates at least in
first order terms.

\subsection{Multipole moments}
Quo-harmonic coordinates are asymptotically Cartesian and mass-centered
coordinates in the sense of Thorne \cite{Thorne1} (see Appendix). Then we
can read the multipole moments of the vacuum exterior field  from the
$g_{tt}$ and $g_{t\varphi}$ components of the metric (\ref{quo2}).
However we do not need to do this search because they must be equal to
the constants of the exterior metric that appear in these two components.
Lichnerowicz's matching provides for them the following values:
\begin{eqnarray}
&&M_0 \approx \lambda r_0\left[1 + \frac{\lambda}{5}\left(21+2\Omega^2\right)\right]\,,
\nonumber\\
&&M_2 \approx -\frac12\lambda\Omega^2r_0^3\left(1 +\frac{139}{35} \lambda\right)\,,
\nonumber\\
&&\ \,
J_1 \approx \frac25\lambda^{3/2}\Omega r_0^2\left(1+\frac56\Omega^2\right)\,,\nonumber\\
&&\ \,
J_3 \approx -\frac17\lambda^{3/2}\Omega^3r_0^4\,.
\label{quomoments}
\end{eqnarray}

These expressions look like those we obtained for the multipole moments in the
harmonic case but  differ from them in $\lambda^2$ terms.
Nevertheless, multipole moments are coordinate independent quantities so
both expressions must lead to the same numerical values for any set of
the source parameters we choose. Let us be more precise. The parameters
$\mu$ and $\omega$ are clearly constants that characterize the fluid but $r_0$, the mean source radius, is a constant of a
different class: it can be defined as the value taken by the radial
coordinate on the matching surface corresponding to the static limit of our metric. Therefore $r_0$ does not have to be the same
in quo-harmonic coordinates and in harmonic coordinates.
However they both must be linked if they parametrize the
same surface of the space-time; that is, if we are talking
about just a single member of this one-parameter family of metrics. We can
obtain some insight into the relationship between quo-harmonic and harmonic
values of
$r_0$ which lead to the same metric by identifying the expressions of
multipole moments in both systems of coordinates. Equation (\ref{quomoments}) can be transformed into (\ref{moments}) by
means of the following substitutions 
\begin{equation}
r_0 \ \longrightarrow\  r_0\left(1-\frac25\lambda\right)\,,\quad
 \lambda \longrightarrow\  \lambda\left(1-\frac45\lambda\right)\,.
\label{quotoh}
\end{equation}
Let us note that $\lambda$ is a function of $r_0$, then it has
different values in one or another system of coordinates this is
the meaning of the second substitution. On the other hand $\Omega$ is
just a function of $\mu$ and $\omega$ so it does not change.

The above comment points to the change from quo-harmonic to harmonic coordinates 
for the Schwarzschild metric to be in the origin of  the substitution rule (\ref{quotoh}). That is true. If we substitute the quo-harmonic radial coordinate in
the static limit of the quo-harmonic metric (\ref{quo2}) up to order $\lambda$,
\begin{eqnarray}
&&
\bm{g}_{\rm int}\approx
\left(-1+3\lambda-\lambda\eta^2\right)\,\bm{T}_0
+\left(1+4\lambda-\frac43\lambda\eta^2\right)\bm{D}_0
+\frac{2}{15}\lambda\eta^2\,\bm{F}_2\,,
\nonumber\\
[.6ex]
&&
\bm{g}_{\rm ext}\approx
\left(-1+2\frac{\lambda}{\eta}\right)\bm{T}_0
+\left(1+8\frac{\lambda}{3\eta}\right)\bm{D}_0
+\frac{\lambda}{3\eta}\left(1-\frac{3}{5\eta^2}\right)\bm{F}_2\,,
\end{eqnarray}
in the following way,
\begin{equation}
r\ \longrightarrow\ \left\{
\begin{array}{ll}
\displaystyle
 r\left[1-\frac{\lambda}{2}\left(1-\frac{r^2}{5r_0^2}\right)\right]\,,
 &{\rm for}\ \bm{g}_{\rm int}\\
 &\\
 \displaystyle
 r\left[1-\lambda\frac{r_0}{2r}\left(1-\frac{r_0^3}{5r^3}\right)\right]\,,
 &{\rm for}\ \bm{g}_{\rm ext}\\
\end{array}
\right.
\end{equation}
which implies that $r_0$ and $\lambda$ have to be substituted at the same time in the way we suggested in (\ref{quotoh}). If so we reach the static limit for the
harmonic case metric (\ref{Schwarz}).

The substitutions (\ref{quotoh}) make no sense when we apply them to 
other exterior or interior constants. In fact, $A_2$, $B_2$,
$B_4$, $a_0$, $b_0$, $a_2$ and $b_2$, which were at least of order
$\lambda^2$ in the harmonic case, are of order $\lambda$ in the
quo-harmonic case. It makes  all the calculations more cumbersome and the
final expression of the metric more involved, as one can see by comparing
(\ref{quo2}) to (\ref{g2}).

\section*{Acknowledgments}

This study was supported partly by Projects BFM2003--02121 and BFM 2003--07076 from 
Ministerio de Educaci\'on y Ciencia and Project SA010C05
from Junta de Castilla y Le\'on. The authors thank J. M.
Aguirregabir\'\i{}a and M. Mars for helpful comments and discussions.

\section*{Appendix: Approximate metric}

Here we write down the components of the metric in the original
Papapetrou form (\ref{eqmetrica}). This is a standard way to introduce
stationary axisymmetric metrics. Therefore, some interesting properties
of the metric structure can be easier recognized looking at the following
expressions than they could be by examining the formulas of the preceding
sections, even though they are more suitable to run the perturbation algorithm.
We use the obvious shorthand notation $P_l^m\equiv P_l^m(\cos\theta)$.

\subsection*{Harmonic coordinates}

{\bf A.1 Interior metric components in harmonic coordinates}

\begin{eqnarray*}
&&\gamma_{tt}=
-1 + 3\lambda +\frac14\lambda^2\left(21 + 2\Omega^2\right)-\lambda\eta^2\left(1+\frac32\lambda-\lambda\Omega^2\right)\\
&&\ \ \quad
+\frac12\lambda^2\eta^4\left(\frac12-\frac75\Omega^2\right)
-\lambda\Omega^2\eta^2\left(1+\frac{73}{70}\lambda-\frac{15}{14}\lambda\eta^2\right)P_2\,,\\
[.6ex]
&&\gamma_{t\varphi}=
 2\lambda^{3/2}\Omega\eta\left(1 + \frac13\Omega^2-\frac35\eta^2\right)P_1^1
- \frac{2}{7}\lambda^{3/2}\Omega^3\eta^3 P_3^1\,,\\
[.6ex]
&&\gamma_{\varphi\varphi}=
1 + 3\lambda+\frac72\lambda^2\left(\frac72 +\frac13\Omega^2\right)
-\lambda\eta^2\left(1+\frac{57}{10}\lambda-\frac{19}{105}\lambda\Omega^2\right)\\
&&\ \ \quad
+\frac{1}{7}\lambda^2\eta^4\left(\frac{19}{4}-\frac{128}{45}\Omega^2\right)
-\lambda\Omega^2\eta^2\left(1+\frac{1079}{210}\lambda-\frac{106}{63}\lambda\eta^2\right)P_2\,,\\
[.6ex]
&&\gamma_{rr}=
1 + 3\lambda +\frac72\lambda^2\left(\frac72 +\frac13\Omega^2\right)
- \lambda\eta^2\left(1+\frac{51}{10}\lambda+\frac35\lambda\Omega^2\right)\\
&&\ \ \quad
+\frac{11}{14}\lambda^2\eta^4\left(\frac12+\frac{\Omega^2}5\right)
-\lambda\Omega^2\eta^2\left(1+\frac{253}{70}\lambda-\frac{17}{42}\lambda\eta^2\right)P_2\,,\\
[.6ex]
&&\gamma_{r\theta}=
\frac13\lambda^2\Omega^2\eta^2\left(\frac47-\frac13\eta^2\right) P_2^1\,,\\
[.6ex]
&&\gamma_{\theta\theta}=
1 + 3\lambda  +\frac72\lambda^2\left(\frac72 +\frac{\Omega^2}3\right)
- \lambda\eta^2\left(1+\frac{57}{10}\lambda+\frac{61}{105}\lambda\Omega^2\right)\\
&&\ \ \quad
+\frac{1}{7}\lambda^2\eta^4\left(\frac{19}{4}+\frac{47}{45}\Omega^2\right)
-\lambda\Omega^2\eta^2\left(1+\frac{919}{210}\lambda-\frac{71}{63}\lambda\eta^2\right)P_2\,.
\end{eqnarray*}

\noindent
{\bf A.2 Exterior metric components in harmonic coordinates}

\begin{eqnarray*}
&&\gamma_{tt}=
 -1+2\frac{\lambda}{\eta}\left[1+\lambda\left(3 + \frac{2}{5}\Omega^2\right)-\frac{\lambda}{\eta}\right]
-\frac{\lambda\Omega^2}{\eta^3}\left(1+\frac{69}{35}\lambda -2\frac{\lambda}{\eta}\right)P_2\,,\\
[.6ex]
&&\gamma_{t\varphi} = 
2\frac{\lambda^{3/2}\Omega}{\eta^2}\left(\frac25+\frac{1}{3}\Omega^2\right)P_1^1
-2\frac{\lambda^{3/2}\Omega^3}{7\eta^4}P_3^1\,,\\
[.6ex]
&&\gamma_{\varphi\varphi}=
1 + 2\frac{\lambda}{\eta}\left[1+\lambda\left(3+\frac{2}{5}\Omega^2\right)+\frac{\lambda}{2\eta}
+2 \frac{\lambda}{35\eta^2}\left(2+\frac13\Omega^2\right)
 \right. \\
&&\ \ \quad
\left.
+\frac{\lambda\Omega^2}{12\eta^3}- 2\frac{\lambda\Omega^2}{63\eta^4}\right] 
-\frac{\lambda\Omega^2}{\eta^3}\left(1+\frac{69}{35}\lambda+7\frac{\lambda}{6\eta}
+20\frac{\lambda}{63\eta^2}\right) P_2\,,\\
[.6ex]
&&\gamma_{rr}=
1+2\frac{\lambda}{\eta} \left[1
+ \lambda\left(3+\frac{2}{5}\Omega^2\right) +\frac{\lambda}{\eta}
-4\frac{\lambda}{35\eta^2}\left(2+\frac{1}3\Omega^2\right) \right]\\
&&\ \ \quad
-\frac{\lambda\Omega^2}{\eta^3} \left(1
+\frac{69}{35}\lambda+2\frac{\lambda}{\eta}-16\frac{\lambda}{21\eta^2}\right) P_2\,,\\
[.6ex]
&&\gamma_{r\theta}=
\frac{\lambda^2\Omega^2}{3\eta^4}\left(1-\frac{16}{21\eta}\right) P_2^1\,,\\
[.6ex]
&&\gamma_{\theta\theta}=
1 + 2\frac{\lambda}{\eta}\left[1+\lambda\left(3+\frac{2}{5}\Omega^2\right)+\frac{\lambda}{2\eta}
+2 \frac{\lambda}{35\eta^2}\left(2+\frac13\Omega^2\right)
\right. \\
&&\ \ \quad
\left.
-\frac{\lambda\Omega^2}{12\eta^3}+ 2\frac{\lambda\Omega^2}{63\eta^4}\right] 
-\frac{\lambda\Omega^2}{\eta^3}\left(1+\frac{69}{35}\lambda+5\frac{\lambda}{6\eta}
+4\frac{\lambda}{9\eta^2}\right) P_2\,,\\
\end{eqnarray*}

\subsection*{Quo-harmonic coordinates}

{\bf A.3 Interior metric components in quo-harmonic coordinates}

\begin{eqnarray*}
&&\hspace*{-1em}\gamma_{tt}=
-1+3\lambda+\frac12\lambda^2\left(\frac{153}{10}+\Omega^2\right)
-\lambda\eta^2\left(1+\frac52\lambda-\lambda\Omega^2\right)\\
&&\quad
+\frac{1}{10}\lambda^2\eta^4\left(\frac92-7\Omega^2\right)
-\lambda\Omega^2\eta^2\left(1+\frac{499}{210}\lambda-\frac{109}{70}\lambda\eta^2\right)P_2\,,\\
[.6ex]
&&\hspace*{-1em}\gamma_{t\varphi}=
2\lambda^{3/2}\Omega\eta\left(1+\frac13\Omega^2-\frac35\eta^2\right)P_1^1
-\frac27\lambda^{3/2}\Omega^3\eta^3P_3^1\,,\\
[.6ex]
&&\hspace*{-1em}\gamma_{\varphi\varphi}=
1+\lambda\left(4-\frac16\Omega^2\right)+\frac{1}{10}\lambda^2\left(223+\frac{13}{14}\Omega^2\right)\\
&&\quad
-\frac15\lambda\eta^2\left(6-\frac{5}{14}\Omega^2+48\lambda-\frac{517}{84}\lambda\Omega^2\right)
+\frac{1}{35}\lambda^2\eta^4\left(\frac{216}{5}-\frac{557}{24}\Omega^2\right)\\
&&\quad
-\frac17\lambda\Omega^2\eta^2\left(8+\frac{185}{3}\lambda-\frac{593}{30}\lambda\eta^2\right)P_2\,,\\
[.6ex]
&&\hspace*{-1em}\gamma_{rr}=
1+4\lambda+\lambda^2\left(\frac{223}{10}+\frac43\Omega^2\right)-\frac85\lambda\eta^2\left(1+8\lambda\right)
+\frac{1}{35}\lambda^2\eta^4\left(\frac{352}{5}-12\Omega^2\right)\\
&&\quad
+\lambda\Omega^2 \left[\frac13+\frac{521}{210}\lambda
-\frac17\eta^2\left(13+\frac{3023}{30}\lambda\right)+\frac{683}{140}\lambda\eta^4 \right]P_2\,,\\
[.6ex]
&&\hspace*{-1em}\gamma_{r\theta}=
\frac12\lambda\Omega^2 \left[ \frac13+\frac{521}{210}\lambda
-\frac17\eta^2\left(3+\frac{169}{6}\lambda\right)+\frac{89}{84}\lambda\eta^4 \right] P^1_2\,,\\
[.6ex]
&&\hspace*{-1em}\gamma_{\theta\theta}=
1+\lambda\left(4+\frac16\Omega^2\right)+\frac{1}{10}\lambda^2\left(223+\frac{1081}{42}\Omega^2\right)\\
&&\quad
-\frac15\lambda\eta^2\left(6+\frac{5}{14}\Omega^2+48\lambda+\frac{517}{84}\lambda\Omega^2\right)
+\frac{1}{35}\lambda^2\eta^4\left(\frac{216}{5}+\frac{173}{24}\Omega^2\right)\\
&&\quad
-\lambda\Omega^2 \left[\frac13+\frac{521}{210}\lambda
+\eta^2\left(1+\frac{1333}{210}\lambda\right)-\frac{821}{420}\lambda\eta^4 \right]P_2\,.
\end{eqnarray*}

\noindent
{\bf A.4 Exterior metric components in quo-harmonic coordinates}

\begin{eqnarray*}
&&\hspace*{-1.5em}\gamma_{tt}=
-1+2\frac{\lambda}{\eta} \left[1+\frac{\lambda}{5}\left(21+2\Omega^2\right)
-3\frac{\lambda}{2\eta}+\frac{\lambda}{10\eta^2} \right]\\
&&
-\frac{\lambda\Omega^2}{\eta^3} \left(1
+\frac{139}{35}\lambda-11\frac{\lambda}{3\eta}+18\frac{\lambda}{35\eta^3}\right) P_2\,,\\
[.6ex]
&&\hspace*{-1.5em}\gamma_{t\varphi}= 
2\frac{\lambda^{3/2} \Omega}{\eta^2} \left(\frac25+\frac13\Omega^2\right) P_1^1
-2\frac{\lambda^{3/2} \Omega^3}{7\eta^4}P_3^1\,,\\
[.6ex]
&&\hspace*{-1.5em}\gamma_{\varphi\varphi}=
1+\frac{\lambda}{\eta} \left[3+\frac35\lambda\left(21+2\Omega^2\right)+9\frac{\lambda}{4\eta}
-\frac{1}{\eta^2}\left(\frac15+\frac16\Omega^2+\frac{31}{35}\lambda+\frac{7}{10}\lambda\Omega^2\right)
\right.\\
&&
\left.
-5\frac{\lambda\Omega^2}{24\eta^3}+\frac{\Omega^2}{14\eta^4}\left(1+\frac{67}{15}\lambda\right)
-\frac{\lambda}{4\eta^5}\left(\frac{3}{25}-\frac17\Omega^2\right)+\frac{\lambda\Omega^2}{70\eta^6} \right]\\
&&
-\frac{\lambda\Omega^2}{\eta^3} \left[\frac32+\frac{417}{70}\lambda+17\frac{\lambda}{12\eta}
-\frac{1}{14\eta^2}\left(5+\frac{67}{3}\lambda\right)+12\frac{\lambda}{35\eta^3}-19\frac{\lambda}{140\eta^5} \right] P_2\,,\\
[.6ex]
&&\hspace*{-1.5em}\gamma_{rr}=
1+2\frac{\lambda}{\eta} \left[ 1+ \frac{\lambda}{5}\left(21+2\Omega^2\right)+\frac{\lambda}{2\eta}
+\frac1{\eta^2} \left(\frac15+\frac{31}{35}\lambda+\frac{2}{21}\lambda\Omega^2\right)
\right.\\
&&
\left.
+\frac{\lambda}{20\eta^3}+3\frac{\lambda}{25\eta^5}  \right]
-\frac{\lambda\Omega^2}{\eta^3}\left[ \frac23+\frac{254}{105}\lambda+5\frac{\lambda}{12\eta}
+\frac{6}{7\eta^2}\left(1+\frac{67}{15}\lambda\right)
\right.\\
&&
\left.
-7\frac{\lambda}{10\eta^3}+15\frac{\lambda}{14\eta^5}\right] P_2\,,\\
[.6ex]
&&\hspace*{-1.5em}\gamma_{r\theta}=
-\frac{\lambda\Omega^2}{\eta^3}\left[ \frac13+\frac{136}{105}\lambda+ 5\frac{\lambda}{24\eta}-\frac{2}{7\eta^2}\left(1+\frac{67}{15}\lambda\right)+5\frac{\lambda}{28\eta^3}-23\frac{\lambda}{140\eta^5} \right] P^1_2\,,\\
[.6ex]
&&\hspace*{-1.5em}\gamma_{\theta\theta}=
1+\frac{\lambda}{\eta} \left[3+\frac35\lambda\left(21+2\Omega^2\right)+9\frac{\lambda}{4\eta}
-\frac{1}{\eta^2}\left(\frac15-\frac16\Omega^2+\frac{31}{35}\lambda-\frac{107}{210}\lambda\Omega^2\right)
\right.\\
&&
\left.
+5\frac{\lambda\Omega^2}{24\eta^3}-\frac{\Omega^2}{14\eta^4}\left(1+\frac{67}{15}\lambda\right)
-\frac{\lambda}{4\eta^5}\left(\frac{3}{25}+\frac17\Omega^2\right)-\frac{\lambda\Omega^2}{70\eta^6} \right]\\
&&
-\frac{\lambda\Omega^2}{2\eta^3} \left[\frac{11}{3}+ \frac{43}{3}\lambda+11\frac{\lambda}{3\eta}
-\frac{1}{\eta^2}\left(1+\frac{67}{15}\lambda\right)+19\frac{\lambda}{35\eta^3}-23\frac{\lambda}{70\eta^5}\right]P_2\,.
\end{eqnarray*}

\end{document}